\documentclass[12pt]{article}
\usepackage{amsmath}
\usepackage{graphicx,psfrag,epsf}
\usepackage{enumerate}
\usepackage{natbib}

\usepackage{url}
\usepackage{algorithm}
\usepackage{algorithmic}

\newcommand{\blind}{1}

\begin{document}

\def\spacingset#1{\renewcommand{\baselinestretch}%
{#1}\small\normalsize} \spacingset{1}


\if1\blind
{
  \title{\bf A Flexible Joint Longitudinal-Survival Model for Analysis of End-Stage Renal Disease Data}
  \author{Sepehr Akhavan Masouleh, Tracy Holsclaw, Babak Shahbaba, \\and Daniel L. Gillen \\
    Department of Statistics, University of California, Irvine}
  \maketitle
} \fi

\if0\blind
{
  \bigskip
  \bigskip
  \bigskip
  \begin{center}
    {\LARGE\bf A Flexible Joint Longitudinal-Survival Model for Analysis of End-Stage Renal Disease Data}
\end{center}
  \medskip
} \fi

\bigskip
\begin{abstract}
{We propose a flexible joint longitudinal-survival framework to examine the association between longitudinally collected biomarkers and a time-to-event endpoint. More specifically, we use our method for analyzing the survival outcome of end-stage renal disease patients with time-varying serum albumin measurements. Our proposed method is robust to common parametric assumptions in that it avoids explicit distributional assumptions on longitudinal measures and allows for subject-specific baseline hazard in the survival component. Fully joint estimation is performed to account for the uncertainty in the estimated longitudinal biomarkers included in the survival model.}
\end{abstract}

\noindent%
\vfill

\newpage
\spacingset{1.45} 

\section{Introduction}
\label{uniJointIntro}

In this paper, we propose a flexible joint longitudinal-survival framework for quantifying the association between longitudinally measured biomarkers (e.g., serum albumin) and time-to-death among end-stage renal disease (ESRD) patients. ESRD is a condition where the filtration performed by the kidneys has been reduced to a point at which life can no longer adequately be sustained. According to data from the National Institute of Diabetes and Digestive and Kidney Diseases (NIDDK) over 850,000 persons in the United States are being treated for ESRD and many more suffer from early stage chronic kidney disease. The standard of care for adult ESRD patients that do not have access to a viable transplant is hemodialysis. Hemodialysis patients experience extremely high mortality rates. Multiple epidemiologic studies have shown that indices of protein-energy malnutrition (PEM) are a strong predictor of total mortality among hemodialysis patients \citep{fung02, wong02}. Serum albumin, a protein biomarker and surrogate for nutritional status, is among the indices of PEM that have been associated with mortality. \cite{fung02} found that among hemodialysis subjects, baseline albumin level and the slope of albumin over time were independent risk factors for mortality, suggesting that the low albumin level is not simply a consequence of co-morbidities associated with dialysis but may be a precursor. This analysis did not consider other potential characteristics of the within-subject changes in serum albumin that may also be associated with mortality in this high risk population. It is natural to hypothesize that high within-subject variability in serum albumin measured over time may also be indicative of increased mortality. That is, non-linear patterns or high instability around a patient's first-order trend is likely an indication of nutritional instability (possibly due to inadequate dialyzing) and hence may be a risk factor for morbidity and mortality. While this association is plausible, it has not been considered in the literature to the best of our knowledge. The reason for this may be due to the difficulty in summarizing and \emph{efficiently} estimating within-subject variability around a specified mean model using standard statistical methodology. However, if the hypothesis were established, this summary measure would provide nephrologists with an additional biomarker to monitor ESRD patients and potentially decrease the risk of mortality in these patients. This is one of the main objectives of our study. To this end, we propose a flexible joint longitudinal-survival model. 

Flexibility in our joint model is achieved in the longitudinal component via the use of a Gaussian process prior with a parameter that captures within-subject volatility in time-varying biomarkers. The survival component of our proposed models quantifies the association between the longitudinally measured biomarkers and the risk of mortality using a Dirichlet process mixture of Weibull distributions. The clustering mechanism of the Dirichlet process provides a framework for borrowing information when estimating subject-specific baseline hazards in the survival component. Estimation for the longitudinal and survival parameters is carried out simultaneously via Bayesian parameter posterior sampling approach. 

In contrast to most competing models, our proposed method provides the following advantages: 1) it avoids relying on restrictive parametric assumptions; 2) it properly accounts for variability in longitudinal measures; 3) it provides a natural framework to capture the association of both first and second moments of the distribution of the longitudinal covariate with survival; and 4) its underlying mechanism for clustering patients can help clinicians to design more personalized treatments. 

In what follows, we provide an overview of some related methods and discuss the specific dataset used in our research.   

\subsection{Some Related Methods}

Survival analysis often involves evaluating effects of longitudinally measured biomarkers on mortality. When longitudinal measures are sparsely collected, incomplete, or prone to measurement error, including them directly as a traditional time-varying covariate in a survival model may lead to biased regression estimates \citep{prentice1982}. To address this issue, one could apply a two-stage method, where the first stage consists of modeling the longitudinal components via a mixed-effects model, and in the second stage, the modeled values or their summaries (e.g., first-order trends) are included in a survival model \citep{dafni1998,tsiatis1995}. Standard approaches for analyzing longitudinal covariates include frequentist mixed effects models \citep{verbeke00, pinheiro00} and Bayesian hierarchical models \citep{christensen10, gelman03}. In general, these approaches parameterize the population mean model as a function of potentially time-varying covariates, subject-specific deviations from the population mean via random effects, and residual variability by subject level variance-covariance matrices that generally account for serial auto-correlation. More recently, several authors have proposed Gaussian process (GP) models as an alternative to more standard longitudinal regression methods since they easily allow for flexible model specification in a coherent probabilistic framework \citep{shi07, liu10}.

Two-stage methods in general fail to account for uncertainty in the estimated longitudinal summary measures. To overcome this issue, several joint longitudinal-survival models have been proposed \citep{prentice1982,bycott1998,hanson2011,wang2001,faucett1996,brown2003,wulfsohn1997,song2002,law2002}. These models account for uncertainty in longitudinal measures by modeling them simultaneously with the survival outcome. However, most existing joint models still rely on multiple restrictive parametric and semi-parametric assumptions and generally focus only on associating the first moment of the distribution of the longitudinal covariate with survival. In this paper, we address these issues by developing a flexible joint longitudinal-survival framework that avoids simple distributional assumptions on longitudinal measures and allow for subject-specific baseline hazard in modeling the survival outcome.

\subsection{United States Renal Data System}

The specific dataset we consider in this study is obtained from the United States Renal Data System (USRDS). The USRDS collects and maintains demographic, treatment, and mortality on all Medicare covered ESRD patients undergoing renal replacement therapy. Details of the USRDS are described elsewhere \citep{usrds}. In addition to its standard data collection, the USRDS has periodically conducted individual special studies. Termed the Dialysis Morbidity and Mortality Studies (DMMS), these specialized studies were carried out to obtain more specific information on smaller samples of ESRD patients. Four DMMS studies were conducted in waves in the 1990s. The first of these studies, DMMS Wave 1 (DMMS-1), included a nutritional sub-cohort of $N$=2,613 hemodialysis patients in which monthly and bimonthly PEM index measures, including serum albumin, were longitudinally collected over 18 months. Given the longitudinal nature of DMMS-1, the study is ideal for estimating within-subject changes in serum albumin. In addition, because the USRDS collects and maintains data on the time and cause of death, this study can also be used to assess whether derived characteristics of within-subject changes in serum albumin are associated with mortality.

\section{Methodology}\label{UniJointMethod}
In this section, we provide the details of our proposed joint models for a longitudinal covariate, $\boldsymbol{X}$, and a survival outcome, $\boldsymbol{Y}$. Throughout this section, we consider $n$ independent subjects where $l_i$ longitudinal measurements,  $X_{ij}$, are obtained for subject $i$ at time points $t_{ij}$, $j = 1, \dots, l_i$. Also, associated with each subject, there is an observed survival time, $Y_i \equiv \mbox{min}\{T_i, C_i\}$ and event indicator $\delta_i\equiv 1_{[Y_i=E_i]}$, where $T_i$ and $C_i$ denote the true event and censoring time for subject $i$, respectively. Further, we make the common assumption that $C_i$ is independent of $T_i$ for all $i$, $i=1,\dots,n$.

\subsection{The Joint Model}\label{M_Joint}
Being interested in estimating the effect of longitudinal measures on survival outcomes, for specifying the joint model likelihood we took a similar approach as \cite{brown2003}, where we define the contribution of each subject to the joint model likelihood as the multiplication of the likelihood function of the longitudinal measures for that subject and her/his time-to-event likelihood that is conditioned on her/his longitudinal measures. Let $f^{(i)}_L$, $f^{(i)}_{S|L}$, and $f^{(i)}_{L, S}$ denote the longitudinal likelihood contribution, the conditional survival likelihood contribution, and the joint likelihood contribution for subject $i$. One can write the joint longitudinal-survival likelihood function as
\begin{eqnarray} 
f_{L, S} &=& \prod_{i=1}^{n}f^{(i)}_{L, S} = \prod_{i=1}^{n} \big(f^{(i)}_L \times f^{(i)}_{S|L}\big).\label{jointMLikeLiHood} 
\end{eqnarray}

\subsection{Longitudinal Component}\label{M_Long}
We motivate the development of the Gaussian process model for the longitudinal biomarker by first considering the following simple linear model for estimating the trend in the biomarker for a single subject $i$ with an $l_i \times 1$ vector of measure biomarkers of $\boldsymbol{X_i}$ which is of the form

\begin{eqnarray*}
\boldsymbol{X_i} = 
 \begin{pmatrix}
  X_i(t_{i1})\\
  X_i(t_{i2})\\
  \vdots\\
  X_i(t_{il_i})
 \end{pmatrix},
\end{eqnarray*}
where
\begin{eqnarray*} 
\boldsymbol{X_i} | \beta^{(L)}_{i0}  \sim N(\boldsymbol{\beta^{(L)}_{i0}},\boldsymbol{\Sigma_i}).
\end{eqnarray*}
with $\beta^{(L)}_{i0}$ as the subject-specific intercept, $\boldsymbol{\beta^{(L)}_{i0}}$ is vector of repeated $\beta^{(L)}_{i0}$ value that is of size $l_i \times 1$, and $\boldsymbol{\Sigma_i} = \sigma^2 I_{l_i \times l_i}$. 

By adding a stochastic component that is indexed by time in the model, one can extend the model to capture non-linear patterns over time. Specifically, we consider a stochastic vector, $\boldsymbol{W}$, that is a realization from a Gaussian process prior, $W(t)$ with mean zero and covariance function $C(t,t')$.  Thus for subject $i$, $\boldsymbol{W_i}  \sim N_{l_i}(\boldsymbol{0}, \boldsymbol{C_{l_i \times l_i}})$, where $\boldsymbol{W_i} = (W_{t_{i1}}, \dots, W_{t_{il_i}})'$ and the $(j,j')$ element of $\boldsymbol{C_{l_i \times l_i}}$ is given by $C(t_{ij},t_{ij'})$,$j , j' \in \{1, \dots, l_i\}$. We characterize the covariance function, $\boldsymbol{C_{l_i \times l_i}}$, using the following squared exponential form

\begin{eqnarray*}
\boldsymbol{C_{l_i \times l_i}}(j,j') = {\kappa_i}^2 e^{-\rho^2 (t_{ij} - t_{ij'})^2}.
\end{eqnarray*}

\begin{figure}
\centering\includegraphics[scale = 0.8]{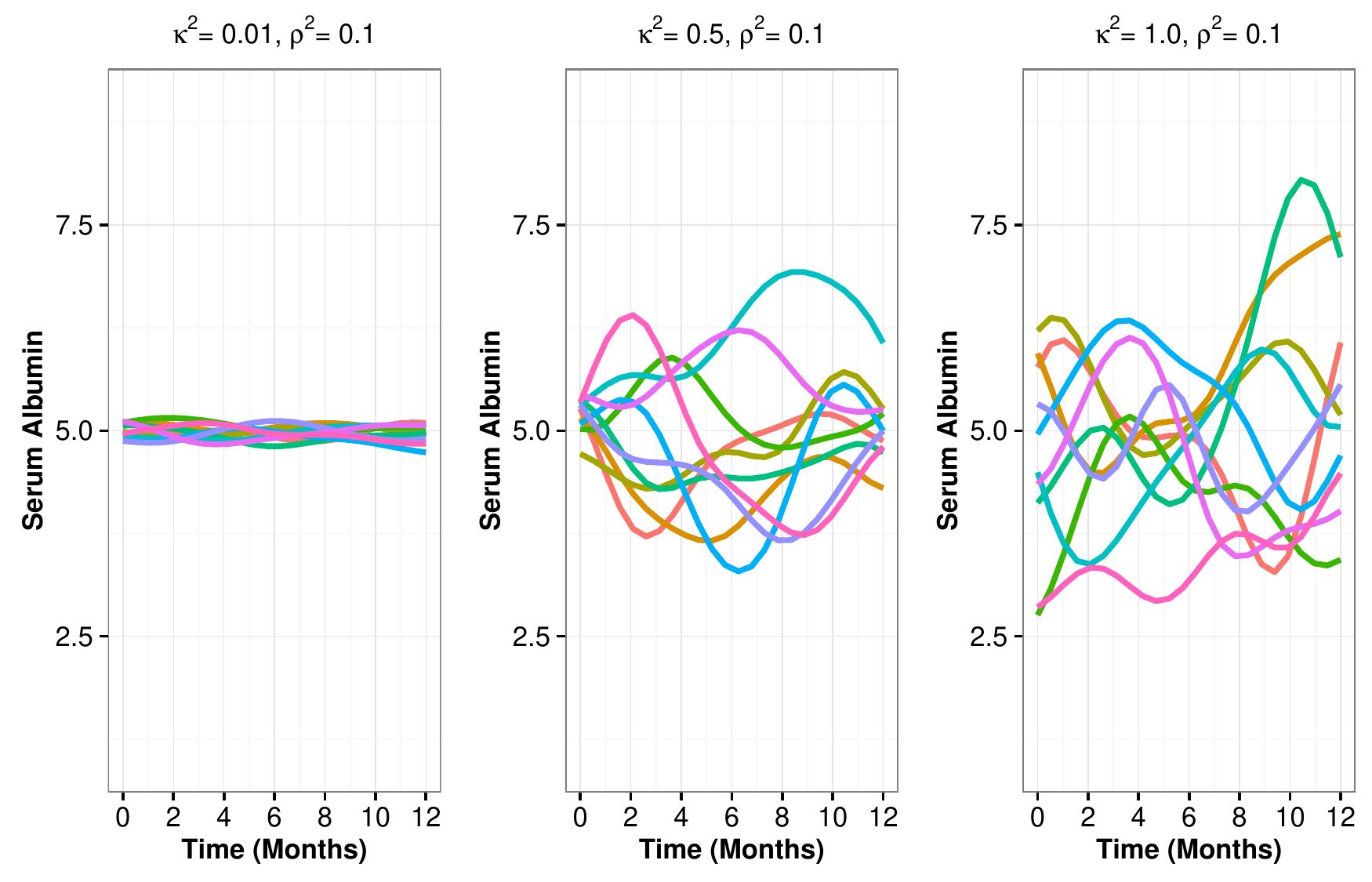}
\caption{With a fixed correlation length parameter $\rho^2$, $\kappa^2$ parameter captures volatility in Gaussian process models with the squared exponential covariance function. In each plot, ten random realizations of the Gaussian process were selected, where the plot to the left has a $\kappa^2$ parameter of 0.01, the plot in the middle has a $\kappa^2$ value of 0.5, and the plot to the right has a $\kappa^2$ value of 1.0. In all plots, correlation length $\rho^2$ is fixed to 0.1 .}
\label{Kappa2SimPlot}
\end{figure}

In this setting, the hyperparameter $\rho^2$ controls the correlation length, and $\kappa^2$ controls the height of oscillations \citep{banerjee2014hierarchical}, and $t_{ij}$ and $t_{ij'}$ are two different time points. For notational simplicity, we define $\boldsymbol{K_i} = e^{-\rho^2 (t_{ij} - t_{ij'})^2} \text{ ;   } j , j' \in \{1, \dots, l_i\}$, and re-write our longitudinal model as 
\begin{eqnarray*}
\boldsymbol{X_i} | \beta^{(L)}_{i0}, {\kappa_i}^2, \rho^2, \sigma^2  \sim N(\boldsymbol{\beta^{(L)}_{i0}}, {\kappa_i}^2 \boldsymbol{K_i} +  \sigma^2 I_{l_i \times l_i}), 
\end{eqnarray*}
where $\sigma^2$  is assumed to be common across all subjects. The correlation length parameter $\rho^2$ controls the maximum distance in time between two time-dependent measurements to be still correlated. This distance for GP models is often called the practical range. \cite{diggle2007springer} defined the practical range for GP as the distance in time between two time-dependent measurements where the correlation between those two measurements is 0.05. With the squared exponential covariance function, that practical range distance is of the form $\sqrt{3/\rho^2}$. At a $\rho^2 = 0.1$, the practical range distance is 5.7 months which is a reasonable range for the real data on end-stage renal disease patients that was obtained from the USRDS. Hence, we fix $\rho^2$ to 0.1, where this value was obtained from the real data on end stage renal disease patients data. By defining our model in this way, subject-specific parameter ${\kappa^2_i}$ will have the role of capturing within-subject volatility of the longitudinal measures. In the context of the motivating USRDS example, ${\kappa^2_i}$ can be of primary scientific interest as it reflects the within-subject volatility (Figure \ref{Kappa2SimPlot}) in serum albumin over time, which is hypothesized to be negatively correlated with longer survival time (Holsclaw et al, 2014). 

We specify the longitudinal component of our joint model to have a likelihood of the form
\begin{eqnarray}
\boldsymbol{X_i} | \boldsymbol{W_i}, \beta^{(L)}_{i0}, {\kappa_i}^2, \rho^2, \sigma^2  & \sim & N(\boldsymbol{\beta^{(L)}_{i0}} + \boldsymbol{W_i}, \sigma^2 I_{l_i \times l_i}),\label{uniJointLong1} 
\end{eqnarray}
where $\boldsymbol{X_i}$ is a vector of longitudinal measures on subject $i$, $\boldsymbol{W_i}$ is a Gaussian process stochastic vector, $\beta^{(L)}_{i0}$ is subject specific intercept for subject $i$, ${\kappa_i}^2$ is a subject-specific measure of volatility for subject $i$, $\rho^2$ is a fixed correlation length, $\sigma^2$ is a measurement error that is shared across all subjects, and finally $I_{l_i \times l_i}$ represents the identity matrix of size $l_i$ where $l_i$ is the number of longitudinal measures on subject $i$. The Gaussian process stochastic vector $\boldsymbol{W_i}$ is distributed Gaussian process as 
\begin{eqnarray}
\boldsymbol{W_i} | {\kappa_i}^2, \boldsymbol{t_i} & \overset{ind}{\sim} & GP_{m_i}(\vec{0}, {\kappa_i}^2\boldsymbol{K_i}),\label{uniJointLong2} 
\end{eqnarray}
where $\boldsymbol{t_i}$ is a vector of the time points at which longitudinal measures on subject $i$ were collected and $\boldsymbol{K_i} = e^{-\rho^2 (t_{ij} - t_{ij'})^2}$, with $t_{ij}$ and $t_{ij'}$ are the $j^{th}$ and ${j'}^{th}$ element of the time vector $\boldsymbol{t_i}$. We assume a Normal prior on the subject-specific random intercepts $\beta^{(L)}_{i0}$ that is of the form
\begin{eqnarray}
\beta^{(L)}_{i0} & \overset{i.i.d}{\sim} & N(\mu_{\beta^{(L)}_{0}}, \sigma^2_{\beta^L_{0}}),
\end{eqnarray}
where $\mu_{\beta^{(L)}_{0}}$ and $\sigma^2_{\beta^L_{0}}$ are prior mean and prior variance respectively. $\boldsymbol{K_i}$, where $i \in \{1, \dots, n\}$ with $n$ as the number of subjects in the study, are assumed to have a log-Normal prior with the prior mean $\mu_{\kappa^2}$ and the prior variance $\sigma_{\kappa^2}$ that is of the form
\begin{eqnarray}
{\kappa_i}^2 & \overset{i.i.d}{\sim} & \textrm{log-Normal}(\mu_{\kappa^2}, \sigma^{2}_{\kappa^2}).
\end{eqnarray}
The correlation length $\rho^2$ is assumed to be fixed and known in our model. Finally, the measurement error $\sigma^2$ is assumed to have a log-Normal prior of the form 
\begin{eqnarray}
\sigma^2 & \sim & \textrm{log-Normal}(\mu_{\sigma^2}, \sigma^{2}_{\sigma^2}),
\end{eqnarray}
where $\mu_{\sigma^2}$ and $\sigma_{\sigma^2}$ are the prior mean and the prior variance respectively.

\subsection{Survival Component}\label{M_Surv}
In order to quantify the association between a longitudinal biomarker and a time-to-event outcome, we define our survival component by using a multiplicative hazard model with the general form of
\begin{eqnarray}
\lambda(T_i | \boldsymbol{{Z_i}^{(s)}}, \boldsymbol{{Z_i}^{(L)}}) = \lambda_0(T_i) exp\{\boldsymbol{\zeta^{(s)}} \boldsymbol{{Z_i}^{(s)}} + \boldsymbol{\zeta^{(L)}} \boldsymbol{{Z_i}^{(L)}(t)}\},
\end{eqnarray}
where $\boldsymbol{{Z_i}^{(s)}}$ is a vector of baseline covariates, $\boldsymbol{{Z_i}^{(L)}} $ is a vector of longitudinal covariates from the longitudinal component of the model, $\lambda_0(T_i)$ denotes the baseline hazard function, and $\boldsymbol{\zeta^{(S)}}$ and $\boldsymbol{\zeta^{(L)}}$ are regression coefficients for the baseline survival covariates and the longitudinal covariates, respectively.

We consider a Weibull distribution for the survival component to allow for log-linear changes in the baseline hazard function over time.  Thus we assume
\begin{eqnarray}
T_i & \sim & \textrm{Weibull}(\tau, \lambda_i),
\end{eqnarray}
where $T_i$ is the survival time, $\tau$ is the shape parameter of the Weibull distribution, and $exp\{\lambda_i\}$ is the the scale parameter of the Weibull distribution. One can write the density function for the Weibull distribution above for the random variable $T_i$ as

\begin{eqnarray}\label{eq:WeibDist}
f(T_i | \tau, \lambda_i) & = & \tau {T_i}^{\tau - 1} exp\big(\lambda_i - exp(\lambda_i) {T_i}^{\tau}\big).
\end{eqnarray}

In this case, the Weibull distribution is available in closed form providing greater computational efficiency.  Under this parameterization, covariates can be incorporated into the model by defining $
\lambda_i = \boldsymbol{\zeta^{(s)}} \boldsymbol{{Z_i}^{(s)}} + \boldsymbol{\zeta^{(L)}} \boldsymbol{{Z_i}^{(L)}}$. In particular, we specify our model as  
\begin{eqnarray}\label{eq:survModel}
T_i | \tau, \boldsymbol{\zeta^{(s)}},  \boldsymbol{\zeta^{(L)}}, \boldsymbol{{Z_i}^{(s)}}, \boldsymbol{{Z_i}^{(L)}} & \sim & \textrm{Weibull}(\tau, \lambda_i = \beta^{(s)}_{i0} + \boldsymbol{\zeta^{(s)}} \boldsymbol{{Z_i}^{(s)}} + \boldsymbol{\zeta^{(L)}} \boldsymbol{{Z_i}^{(L)}}),
\end{eqnarray}
where $\tau$ is a common shape parameter shared across all subjects. $\beta^{(s)}_{i0}$ is a subject specific coefficient in the model which allows for a subject-specific baseline hazard. $\boldsymbol{{Z_i}^{(s)}}$ and $\boldsymbol{\zeta^{(s)}}$ are baseline covariates and their corresponding coefficients, respectively. Finally, $\boldsymbol{{Z_i}^{(L)}}$ and $\boldsymbol{\zeta^{(L)}}$  are coefficients linking the longitudinal parameters of interest to the hazard for mortality. 

In order to avoid an explicit distributional assumption for the survival times, we specify our survival model as an infinite mixture of Weibull distributions that is mixed on the $\beta^{(s)}_{i0}$ parameter. In particular, we use the Dirichlet process mixture of Weibull distributions that is defined as
\begin{eqnarray}\label{eq:SurvDPMprior}
\beta^{(s)}_{i0} | \mu_i, \sigma^2_{\beta^{(s)}_0} &\sim&  N(\mu_i, \sigma^2_{\beta^{(s)}_0}), \\
\mu_i | G &\sim&  G, \\
G &\sim& DP\big(\alpha^{(S)}, G_0),
\end{eqnarray}
where $\sigma^2_{\beta^{(s)}_0}$ is a fixed parameter, $\mu_i$ is a subject-specific mean parameter from a distribution $G$ with a DP prior, $\alpha^{(S)}$ is the concentration parameter of the DP and $G_0$ is the base distribution. By using the Dirichlet process prior on the distribution of $\beta^{(s)}_{i0}$, we allow patients with similar baseline hazards to cluster together which subsequently provides a stronger likelihood to estimate the baseline hazards. For other covariates in the model, we assume a multivariate normal prior of the form
\begin{eqnarray*}
\big( \boldsymbol{\zeta^{(s)}}, \boldsymbol{\zeta^{(L)}} \big) & \sim & MVN(\boldsymbol{0}, {\sigma_0}^2 I),
\end{eqnarray*}
where ${\sigma_0}^2$ is a prior variance and $I$ is an identity matrix.

The shared scale parameter $\tau$ is considered to have a Log-Normal prior of the form 
\begin{eqnarray*}
\tau \sim \textrm{log-Normal}(a_{\tau}, b_{\tau}),
\end{eqnarray*}
where $a_{\tau}$ and $b_{\tau}$ are fixed prior location and prior scale parameters, respectively. 

Finally, we assume that information about the concentration parameter of the Dirichlet process can be specified with the prior
\begin{eqnarray*}
\alpha^{(S)}  \sim  \Gamma(a_{\alpha}^{(S)}, b_{\alpha}^{(S)}),
\end{eqnarray*}
where $a_{\alpha}^{(S)}$ and $b_{\alpha}^{(S)}$ are fixed prior shape and prior scale parameters, respectively.

\subsection{Linking the Two Components}
The proposed modeling framework easily allows for associating multiple summaries of the longitudinal biomarker with the time-to-event outcome. Here we consider three alternative models that incorporate various summary measures of the longitudinal trajectory that are easily and flexibly estimated using the GP model presented in Section \ref{M_Long}:

\begin{itemize}
	\item[] \textbf{Model I}: directly modeling longitudinal outcome at each event time $t$ as a covariate in the survival model:
    $${Z_i}^{(L)} = X^i(t)$$
    
  \item[] \textbf{Model II}: modeling both the value of the longitudinal covariate and also the average rate at which the biomarker changes for each subject. We define this average rate as a weighted area under the derivative curve of the biomarker trajectory
    $$\boldsymbol{{Z_i}^{(L)}} = \big(X_i(t), {X'_{AUC}}^{\tau_0-\tau_1}\big)$$
    $$ \text{where, } {X'_{AUC}}^{\tau_0-\tau_1} = \int_{\tau_0}^{\tau_1} Q(u)X'(u)du$$
  where ${X'_{AUC}}^{\tau_0-\tau_1}$ is a time-dependent covariate that is a weighted average of the derivative of the biomarker trajectory, that is denoted by $X'(u)$ from $\tau_0$ to $\tau_1$ where $\tau_1$ is the time of death for each subject. This average area under the derivative curve can be a weighted average with weights $Q(u)$.
  
  \item[] \textbf{Model III}: modeling summary measures of the longitudinal trajectory. Motivated by the scientific question of interest, in this paper we consider \textbf{random intercepts} and \textbf{subject-specific volatility} as summary measures of interest: $$\boldsymbol{{Z_i}^{(L)}} = \big(\beta^{(L)}_{0i},\kappa^2_i \big)$$
\end{itemize}

Below, we will explain these three models in more detail.
\subsubsection{Model I: a Survival model with Longitudinal Biomarker at event time as a covariate}
This model quantifies the association between a longitudinal biomarker of interest and the time-to-event outcome by directly adjusting for the biomarker measured values in the survival component. While usually biomarkers are measured on a discrete lab-visit basis, the event of interest happens on a continuous basis. While common frequentist models use the so-called last-observation-carried forward (LOCF) technique where the biomarker value at each even time is assumed to be the same as the last measured value for that biomarker, our joint flexible longitudinal-survival model provides a proper imputation method for the biomarker values at each individual's event time. In particular, in each iteration of the MCMC, given the sampled parameters for each individual and by using the flexible Gaussian process prior, there exists posterior trajectories of biomarker for that individual. Our method, then, considers the posterior mean of those trajectories as the proposed trajectory for that individual's biomarker values over time at that iteration. The trajectory, then, can be used to impute time-dependent biomarker covariate value inside the survival component. To be more specific, consider the longitudinal biomarker $\boldsymbol{X_i}$ of the form
\begin{eqnarray*}
\boldsymbol{X_i} | \beta^{(L)}_{i0}, {\kappa_i}^2, \rho^2, \sigma^2  & \sim & N(\boldsymbol{\beta^{(L)}_{i0}}, {\kappa_i}^2\boldsymbol{K_i} + \sigma^2 I_{l_i \times \boldsymbol{l_i}}),
\end{eqnarray*}
where $\beta^{(L)}_{i0}$ is subject-specific random intercept for subject $i$, $\boldsymbol{\beta^{(L)}_{i0}}$ is a vector of repeated subject-specific intercept $\beta^{(L)}_{i0}$ that is of size $l_i \times 1$, ${\kappa_i}^2$ is subject-specific measure of volatility in the longitudinal biomarker for individual $i$, $\rho^2$ is a fixed measure of correlation length, $\sigma^2$ is the measurement error shared across all subjects, $\boldsymbol{K_i}$ is a an $l_i \times l_i$ matrix with it's $jj'$ element as $\boldsymbol{K_i}_{jj'} = e^{-\rho^2 (t_{ij} - t_{ij'})^2}$ where $l_i$ is the number of longitudinal biomarker measures on subject $i$, and $I_{l_i \times l_i}$ is the identity matrix.

For a new time-point $t^{*}$, predicted albumin biomarker for individual $i$ is $X^{*}$ and can be written as
\begin{eqnarray*}
X^{*} | \boldsymbol{X_i}, \boldsymbol{t}, t^{*}  & \sim & N(\mu^{*}, \Sigma^{*}),
\end{eqnarray*}
where the conditional posterior mean $\mu^{*}$ is
\begin{eqnarray}\label{PostMeanPaper1}
\mu^{*} = \beta^L_{i0} + K(t^*, \boldsymbol{t}) {K_{X}}^{-1} (\boldsymbol{X_i} - \beta^{(L)}_{i0}),
\end{eqnarray}
and the conditional posterior variance $\Sigma^{*}$ is
\begin{eqnarray}\label{eq:PostVar}
\Sigma^{*} = K(t^*, t^*) - K(t^*, \boldsymbol{t}) {K_{X}}^{-1} K(t^*, \boldsymbol{t})',
\end{eqnarray}
where $K(t^*, \boldsymbol{t})$ is defined as
\begin{eqnarray}
K(t^*, \boldsymbol{t}) = {\kappa_i}^2 e^{-\rho^2 (t^* - \boldsymbol{t})^2},
\end{eqnarray}
and ${K^{-1}_X}$ is defined as
\begin{eqnarray}
{K^{-1}_X} = (K(\boldsymbol{t}, \boldsymbol{t}) + \sigma^2 I_{l_i \times l_i})^{-1}.
\end{eqnarray}
In order to relate the biomarker value at each time point $t$ to the risk of the event of interest at that time point, "death", we form the survival component of the model as 
\begin{eqnarray*}
T_i | \tau, \boldsymbol{\zeta^{(s)}},  \zeta_{X_{i}}  & \sim & \textrm{Weibull}(\tau, \lambda_i = \beta^{(s)}_{i0} + \boldsymbol{\zeta^{(s)}} \boldsymbol{{Z_i}^{(s)}} + {\zeta_{X_{i}}} X_{i}(t)),
\end{eqnarray*}
where $T_i$ is the survival time, $\tau$ is the shape parameter of the Weibull distribution, $\boldsymbol{\zeta^{(s)}}$ is a vector of coefficients relating baseline survival covariates to the risk of the occurrence of the event of interest, $\zeta_{X_{i}}$ is the coefficient that relates the biomarker value at time $t$ and the risk of "death" at that time point, $\lambda_i$ is the log of the scale parameter in the Weibull distribution, $\beta^{(s)}_{i0}$ is the subject-specific baseline hazard for subject $i$, $\boldsymbol{{Z_i}^{(s)}}$ is a vector of survival coefficients, and $X_{i}(t)$ is the biomarker value at time $t$.

\subsubsection{Model II: A Survival model with covariates of Biomarker Value and the Derivative of its Trajectory at event Time}
In order to get more precision in quantifying the association between the biomarker value at time $t$ and the risk of death, we can extend our proposed model-I by including a measure of the average slope of the biomarker over time. In particular, we define this average slope from time $\tau_0$ to $\tau_1$ as the area under the derivative of the trajectory curve of the biomarker from $\tau_0$ to $\tau_1$. More generally, this area under the curve can be a weighted sum where weights are chosen according to the scientific question of interest. One may hypothesize that the area under the derivative curve that are closer to the event time should be weighted higher compared to the areas that are farther away from time point $t$. In general, we define a weighted area under the derivative curve of the form
\begin{eqnarray*}
{X'_{AUC}}^{\tau_0-\tau_1} = \int_{\tau_0}^{\tau_1} Q(u)X'(u)du,
\end{eqnarray*}
where $\tau_0$ and $\tau_1$ are arbitrary time points chosen according to the scientific question of interest, $Q(t)$ is a weight, and $X'(t)$ represents the derivative of the biomarker over time. In particular, we consider two weighted approaches, where one assumes an equal weight of the form 
\begin{eqnarray*}
Q(t) = \frac{1}{\tau_1 - \tau_0},
\end{eqnarray*}
and another weight of the form 
\[
Q(t)= 
\begin{cases}
  1,& \text{if } t = T_i\\
  0,& \text{otherwise}.
\end{cases}
\]
Under the first weighting scheme, ${X'_{AUC}}$ will be the area under the derivative with equal weights, whereas the second weighting scheme leads to the pointwise derivative value at the the event time. Under this model, the survival component of our joint model will now include two longitudinal covariates, one the biomarker value $X_i(t)$, and another the average derivative of the biomarker trajectory, ${X'_{AUC}}$. 

The derivative of the Gaussian process is still a Gaussian process with the same hyper parameters $\rho^2$ and $\kappa^2$. Therefore, using the same idea of modeling the trajectory of the biomarker, we can also model the derivative of that trajectory. In our model-I, we proposed using the posterior mean of all plausible biomarker trajectories as the proposed trajectory for each subject in order to impute biomarker values at any time point $t$ inside the survival component of the model. Similarly, we propose using the posterior mean of all plausible derivative trajectories for each subject in order to compute the average derivative up until time $t$. Given the fact that differentiation is a linear operation, one can easily compute the posterior mean of the derivative curve by simply switching the order of the differentiation and the expectation as
\begin{eqnarray*}
E(X'_i(t)) & = & E(\frac{\partial X_i(t)}{\partial t}) \\
& = & \frac{\partial \big(E(X_i(t))\big)}{\partial t}.
\end{eqnarray*}
Hence, by using Formula \eqref{PostMeanPaper1} and by taking the derivative of the posterior mean trajectory of the biomarker with respect to time $t^{*}$, the posterior mean of the derivative of the biomarker trajectory is of the form
\begin{eqnarray}\label{eq:DerPostMean}
\frac{\partial \big(E(X_i(t^*))\big)}{\partial t^*} = -2 \rho^2 (t^{*} -  \boldsymbol{t})' \big(K(t^*, \boldsymbol{t}) {K_{X}}^{-1} (\boldsymbol{X_i} - \beta^{(L)}_{i0})\big), 
\end{eqnarray}
where $E\big(X_i(t^*)\big)$ denotes the posterior mean of the biomarker trajectory as a function of time $t^{*}$, $\rho^2$ is the correlation length, $t^*$ is the time-point at which we desire to impute the biomarker value and the average derivative of the biomarker trajectory, $\beta^{(L)}_{i0}$ is subject-specific random intercept, $K(t^*, \boldsymbol{t})$ is defined as
\begin{eqnarray}
K(t^*, \boldsymbol{t}) = {\kappa_i}^2 e^{-\rho^2 (t^* - \boldsymbol{t})^2},
\end{eqnarray}
and ${K^{-1}_X}$ is defined as
\begin{eqnarray}
{K^{-1}_X} = (K(\boldsymbol{t}, \boldsymbol{t}) + \sigma^2 I_{l_i \times l_i})^{-1}.
\end{eqnarray}
Given the the biomarker value $X_i(t)$ and the average derivative value $X'_{AUC, i}(t)$, the survival component of our proposed joint model is of the from
\begin{eqnarray*}
T_i | \tau, \boldsymbol{\zeta^{(s)}},  \zeta_{X_{i}} , \zeta_{X'_{i}}  & \sim & \textrm{Weibull}(\tau, \lambda_i = \beta^{(s)}_{i0} + \boldsymbol{\zeta^{(s)}} \boldsymbol{{Z_i}^{(s)}} + {\zeta_{X_{i}}} X_{i}(t) +  {\zeta_{X'_{i}}} X'_{AUC, i}(t)),
\end{eqnarray*}
where $T_i$ is the survival time, $\tau$ is the shape parameter of the Weibull distribution, $\boldsymbol{\zeta^{(s)}}$ is a vector of coefficients relating baseline survival covariates to the risk of the occurrence of the event of interest, $\zeta_{X_{i}}$ is the coefficient that relates the biomarker value at time $t$ and the risk of "death" at that time point, $\zeta_{X'_{i}}$ is the coefficient that relates the average derivative of the biomarker trajectory up until $t$ and the risk of "death" at that time point,
$\lambda_i$ is the log of the scale parameter in the Weibull distribution, $\beta^{(s)}_{i0}$ is the subject-specific baseline hazard for subject $i$, and $\boldsymbol{{Z_i}^{(s)}}$ is a vector of survival coefficients.

\subsubsection{Model III: A Survival Model with Summary Measures of the Longitudinal Curve as Covariates}
One may choose to characterize longitudinal trajectories with summary measures instead of using the actual biomarker value. In specific, the longitudinal model we proposed provides a natural parameter for describing the within-subject volatility. Given the nature of our proposed longitudinal model, one can summarize the longitudinal trajectory of biomarker by using $\beta^{(L)}_{0i}$ as a measure of subject-specific intercept of longitudinal biomarker as well as $\kappa^2_i$ as a measure of volatility of those trajectories. The survival component of the model is then of the form
\begin{eqnarray*}
T_i | \tau, \beta^{(s)}_{i0}, \boldsymbol{\zeta^{(s)}},  {\zeta_{\beta_{0i}}}^{(L)}, {\zeta_{{\kappa_i}^2}}^{(L)} & \sim & \textrm{Weibull}(\tau, \lambda_i = \beta^{(s)}_{i0} + \boldsymbol{\zeta^{(s)}} \boldsymbol{{Z_i}^{(s)}} + {\zeta_{\beta_{0i}}}^{(L)} {\beta_{0i}}^{(L)} +  {\zeta_{{\kappa_i}^2}}^{(L)} {\kappa_{i}}^2),
\end{eqnarray*}
where $T_i$ is the survival time, $\tau$ is the shape parameter of the Weibull distribution, $\boldsymbol{\zeta^{(s)}}$ is a vector of coefficients relating baseline survival covariates to the risk of the occurrence of the event of interest, ${\zeta_{\beta_{0i}}}^{(L)}$ is the coefficient that relates the subject specific random intercept ${\beta_{0i}}^{(L)}$ and the risk of "death", ${\zeta_{{\kappa_i}^2}}^{(L)}$ is the coefficient that relates the subject-specific measure of volatility of the biomarker measure and the risk of "death", $\lambda_i$ is the log of the scale parameter in the Weibull distribution, $\beta^{(s)}_{i0}$ is the subject-specific baseline hazard for subject $i$, and $\boldsymbol{{Z_i}^{(s)}}$ is a vector of survival coefficients.

\section{Posterior Distribution}
Consider the joint longitudinal-survival likelihood function, $f_{\boldsymbol{L}, \boldsymbol{S}}$, introduced in equation \ref{jointMLikeLiHood}. Let $\boldsymbol{\omega}$ be a vector of all model parameters with the joint prior distribution $\pi(\boldsymbol{\omega})$. The posterior distribution of the parameter vector $\boldsymbol{\omega}$ can be written as
\begin{eqnarray}
\pi(\boldsymbol{\omega} | \boldsymbol{X}, \boldsymbol{Y}) &\propto& f_{L, S} \times \pi(\boldsymbol{\omega}), \label{jointPostDist}
\end{eqnarray}
where $\boldsymbol{X}$ and $\boldsymbol{Y}$ denote longitudinal and time-to-event data respectively, and $f_{L,S}$ is the joint model likelihood function (equation \ref{jointMLikeLiHood}).

The posterior distribution of the parameters in our proposed joint model is not available in closed form. Hence, samples from the posterior distribution of the model parameters are obtained via Markov Chain Monte Carlo (MCMC) methods. We use a hybrid sampling technique where in each iteration of the MCMC, we first sample subject-specific frailty terms in the survival model using Neal's algorithm 8. Then given the sampled frailty terms, we use the Hamiltonian Monte Carlo \citep{HMC} to draw samples from the posterior distribution. Prior distributions on parameters of the joint model were explained in details in Sections \ref{M_Long} and \ref{M_Surv}, and we assume independence among model parameters in the prior (ie. $\pi(\boldsymbol{\omega})$ is the product of the prior components specified previously). We provide further detail on less standard techniques for sampling from the posterior distribution when using a GP prior and issues in evaluating the survival portion of the likelihood function when time-varying covariates are incorporated into the model. 

\subsection{Evaluation of the Longitudinal Likelihood}\label{fastUniGP}
The longitudinal component of our model uses the Gaussian process technique. Gaussian process models are typically computationally challenging to fit because in each iteration of the MCMC the evaluation of the log-posterior probability becomes computationally challenging as the number of measurements increases. In particular, consider our proposed longitudinal model introduced in Section \ref{M_Long} where
\begin{eqnarray}
\boldsymbol{X_i} | \boldsymbol{W_i}, \beta^{(L)}_{i0}, {\kappa_i}^2, \rho^2, \sigma^2  & \sim & N(\boldsymbol{\beta^{(L)}_{i0}} + \boldsymbol{W_i}, \sigma^2 I_{l_i \times l_i}),\nonumber \\
\boldsymbol{W_i} | {\kappa_i}^2, \boldsymbol{t_i} & \sim & GP_{m_i}(\vec{0}, {\kappa_i}^2\boldsymbol{K_i}),\nonumber
\end{eqnarray}
with $\boldsymbol{X_i}$ denoting a vector of longitudinal measures on subject $i$, $\boldsymbol{W}_i$ a Gaussian process stochastic vector, $\beta^{(L)}_{i0}$ a subject specific intercept for subject $i$, ${\kappa_i}^2$ a subject-specific measure of volatility for subject $i$, $\rho^2$ a fixed correlation length, $\sigma^2$ a measurement error that is shared across all subjects, $I_{l_i \times l_i}$ denoting the identity matrix of size $l_i$ with $l_i$ the number of longitudinal measures on subject $i$, $\boldsymbol{t_i}$ a vector of the time points at which longitudinal measures on subject $i$ were collected, and $\boldsymbol{K_i} = e^{-\rho^2 (t_{ij} - t_{ij'})^2}$, where $t_{ij}$ and $t_{ij'}$ are the $j^{th}$ and ${j'}^{th}$ element of the time vector $\boldsymbol{t_i}$. 

In order to sample from the posterior distribution of ${\kappa_i}^2$ and $\sigma^2$ parameters, one can consider a marginal distribution of the following form
\begin{eqnarray}\label{uniGPlongMargLike}
\boldsymbol{X_i} | \beta^{(L)}_{i0}, {\kappa_i}^2, \rho^2, \sigma^2  & \sim & N(\beta^{(L)}_{i0}, {\kappa_i}^2\boldsymbol{K_i} + \sigma^2 I_{l_i \times l_i})
\end{eqnarray}
The marginal distribution above has log-density of the form
\begin{eqnarray}\label{uniGPlongMargLikeDens}
log(f(\boldsymbol{X_i} | \beta^{(L)}_{i0}, {\kappa_i}^2, \rho^2, \sigma^2)) &=& \text{constant} \nonumber\\
&-& \frac{1}{2}log|{\kappa_i}^2\boldsymbol{K_i} + \sigma^2 I_{l_i \times l_i}| \\ 
&-& \frac{1}{2} (\boldsymbol{X_i} - \beta^{(L)}_{i0})^{T} ({\kappa_i}^2\boldsymbol{K_i} + \sigma^2 I_{l_i \times l_i})^{-1}(\boldsymbol{X_i} - \beta^{(L)}_{i0}),\nonumber
\end{eqnarray}
that is the log contribution of subject $i$ to the longitudinal likelihood ( i.e. $log(f^{(i)}_L)$ ). 

sampling from the posterior distribution of ${\kappa_i}^2$ and $\sigma^2$ requires evaluation of the log-density in equation (\ref{uniGPlongMargLikeDens}) that involves evaluation of the determinant and the computation of the inverse of the covariance matrix at each iteration of the MCMC. This process requires $O({l_i}^2)$ memory space and a computation time of $O({l_i}^3)$ per subject, with $l_i$ as the number of within subject measurements. 

In our model setting, we defined $\boldsymbol{K_i} = e^{-\rho^2 (t_{ij} - t_{ij'})}$ with a fixed $\rho^2$ parameter. This means $K_i$ can be pre-computed before starting posterior sampling using MCMC. Furthermore, we propose using the eigenvalue decomposition technique for a faster log-posterior probability computation. Our proposed method was motivated by \cite{flaxman2015fast} and is as follows.

Consider the covariance matrix ${\kappa_i}^2\boldsymbol{K_i} + \sigma^2 I_{l_i \times l_i}$ in the marginal log-density in equation (\ref{uniGPlongMargLikeDens}). Our goal is now to propose a method that makes computation of the inverse and the determinant of this covariance function as efficient as possible. As shown earlier, $\boldsymbol{K_i}$ can be pre-computed before starting the MCMC process as it does not involve any parameter. Consider the eigenvalue decomposition of $\boldsymbol{K_i} = Q \Lambda Q^{T}$, where $\Lambda$ is a diagonal matrix with the eigenvalues of $\boldsymbol{K_i}$ as the diagonal elements, and $Q$ is the corresponding matrix of eigenvectors. ${\kappa_i}^2$ is a scalar parameter that is sampled in each iteration of the MCMC. Multiplication of ${\kappa_i}^2$ times the matrix $\boldsymbol{K_i}$ implies the eigenvalues of this matrix will be ${\kappa_i}^2$ times bigger where the eigenvectors remain the same. Hence, we can conclude that the eigenvalue decomposition of the matrix ${\kappa_i}^2\boldsymbol{K_i}$ is of the form  ${\kappa_i}^2\boldsymbol{K_i} = Q ({\kappa_i}^2\Lambda) Q^{T}$, where $Q$ and $\Lambda$ are elements of the eigenvalue decomposition of the pre-computed matrix $\boldsymbol{K_i}$. Given the pre-computed eigenvalue decomposition of the matrix $\boldsymbol{K_i}$, at each iteration of the MCMC, the determinant of the covariance function of the marginal log-density in equation (\ref{uniGPlongMargLikeDens}) can be computed as
\begin{eqnarray}\label{uniGPlongMargLikeDeterminant}
log|{\kappa_i}^2\boldsymbol{K_i} + \sigma^2 I_{l_i \times l_i}| &=& log|Q ({\kappa_i}^2\Lambda) Q^{T} + \sigma^2 I_{l_i \times l_i}| \nonumber \\
&=& log|Q({\kappa_i}^2\Lambda + \sigma^2 I_{l_i \times l_i})Q^{T}| \nonumber \\
&=& log\big(\prod_{k=1}^{l_i}({\kappa_i}^2\lambda_{ik} + \sigma^2)\big) \nonumber \\
&=& \sum_{k=1}^{l_i}log\big({\kappa_i}^2\lambda_{ik} + \sigma^2 \big).
\end{eqnarray}
In equation (\ref{uniGPlongMargLikeDeterminant}), $\lambda_{ik}$'s are pre-computed eigenvalues of the matrix $\boldsymbol{K_i}$ whereas $\kappa_i$ and $\sigma^2$ are parameters sampled at each iteration of the MCMC. 

Similarly and by using the same trick, we can compute the term $(\boldsymbol{X_i} - \beta^{(L)}_{i0})^{T} ({\kappa_i}^2\boldsymbol{K_i} + \sigma^2 I_{l_i \times l_i})(\boldsymbol{X_i} - \beta^{(L)}_{i0})$ in a more computationally efficient as 
{\footnotesize
\begin{eqnarray}\label{uniGPlongMargLikeInvMatrix}
(\boldsymbol{X_i} - \beta^{(L)}_{i0})^{T} ({\kappa_i}^2\boldsymbol{K_i} + \sigma^2 I_{l_i \times l_i})^{-1}(\boldsymbol{X_i} - \beta^{(L)}_{i0}) &=& (\boldsymbol{X_i} - \beta^{(L)}_{i0})^{T} \big(Q({\kappa_i}^2 \Lambda) Q^{T} + \sigma^2 I_{l_i \times l_i}\big)^{-1} (\boldsymbol{X_i} - \beta^{(L)}_{i0}) \nonumber \\
&=& (\boldsymbol{X_i} - \beta^{(L)}_{i0})^{T} \big(Q({\kappa_i}^2 \Lambda + \sigma^2 I_{l_i \times l_i})Q^{T}\big)^{-1} (\boldsymbol{X_i} - \beta^{(L)}_{i0}) \nonumber \\
&=&(\boldsymbol{X_i} - \beta^{(L)}_{i0})^{T} \big(Q({\kappa_i}^2 \Lambda + \sigma^2 I_{l_i \times l_i})^{-1}Q^{T}\big) (\boldsymbol{X_i} - \beta^{(L)}_{i0}).\nonumber \\
\end{eqnarray}
}
In equation (\ref{uniGPlongMargLikeInvMatrix}), $\boldsymbol{X_i}$ is the data matrix and is fixed, $Q$ and $\Lambda$ are pre-computed eigenvector and diagonal eigenvalue matrices corresponding to the eigenvalue decomposition of the matrix $\boldsymbol{K_i}$. Finally, by utilizing an eigenvalue decomposition, instead of evaluating the term $({\kappa_i}^2\boldsymbol{K_i} + \sigma^2 I_{l_i \times l_i})^{-1}$, one can simply evaluate $\big(Q({\kappa_i}^2 \Lambda + \sigma^2 I_{l_i \times l_i})^{-1}Q^{T}\big)$, where the term $({\kappa_i}^2 \Lambda + \sigma^2 I_{l_i \times l_i})^{-1}$ in the middle is simply the inverse of a diagonal matrix. 

\subsection{Evaluation of the Survival Likelihood}\label{UniGPJointSurvEval}
Here we consider evaluation of the survival component of the decomposed joint likelihood. Consider the survival time for subject $i$ that is denoted by $t_i$ and is distributed according to a Weibull distribution with shape parameter $\tau$ and scale parameter $exp(\lambda_i)$, where $\lambda_i = \boldsymbol{\zeta^{(S)}} \boldsymbol{Z}^{(S)}_i + \boldsymbol{\zeta^{(L)}} \boldsymbol{Z}^{(L)}_i(t)$, where $\boldsymbol{Z}^{(S)}_i$ and $\boldsymbol{Z}^{(L)}_i(t)$ are vectors of covariates for subject $i$, with potentially time-varying covariates, corresponding to the survival and the longitudinal covariates respectively, and $\boldsymbol{\zeta^{(S)}}$ and $\boldsymbol{\zeta^{(L)}}$ are vectors of survival and longitudinal coefficients respectively. One can write the hazard function $h_i(t)$ as
\begin{align}
h_i(t) &= \tau t^{\tau - 1} exp(\lambda_i - exp(\lambda_i) t^\tau). \label{UniGPSurLikeForm}
\end{align}
The survival function $S_i(t)$ can be written as
\begin{align*}
S_i(t) &= exp\{-\int_{0}^{t} h_i(w)dw\}.
\end{align*}
Consider survival data on $n$ subjects, some of whom may have been censored.  Let event indicator $\delta_i$ that is $1$ if the event is observed, and $0$ otherwise. The survival likelihood contribution of subject $i$ can be written in terms of the the hazard function $h_i(t)$ and the survival function $S_i(t)$ as
\begin{align*}
f^{(i)}_{S|L} &=  h_i(t_i)^{\delta_i} S_i(t_i) \\
&=  h_i(t_i)^{\delta_i} e^{-\int_{0}^{t_i}h_i(w)dw}.
\end{align*}
The overall survival log-likelihood can be written as
\begin{align*}
log(L) &= \sum_{i = 1}^{n} log(f^{(i)}_{S|L}) \\
&= \sum_{i = 1}^{n}\big(\delta_i log(h_i(t_i)) - \int_{0}^{t_i}h_i(w)dw\big).
\end{align*}
The hazard function in the equation (\ref{UniGPSurLikeForm}) includes some time-varying covariates which often makes the integral of the hazard function non-tractable. In this case, one can estimate the integral using the rectangular integration as follows:

\begin{algorithm}
\caption*{Integration of Survival Hazard with Time-Varying Covariates}
\begin{algorithmic} 
\STATE 1. Set a fixed number of rectangles $m$ and set $A = 0$
\STATE 2. Divide $(0, t_i)$ interval into $m$ equal pieces each of length $L = t_i/m$
\FOR{$i \in \{1, \dots, m\}$}
\STATE $t_{mid} \leftarrow L/2 + (i - 1)*L$
\STATE $A_{temp} \leftarrow L*h_i(t_{mid})$
\STATE $A \leftarrow A + A_{temp}$
\ENDFOR
\end{algorithmic}
\end{algorithm}

\section{Simulation Studies}
\label{UniJointSim}

In this section, we evaluate our proposed models using a simulation study. We simulated 200 datasets that resembled the real data on end stage renal disease patients that was obtained from the United States Renal Data System (USRDS). To this end, we first simulated longitudinal trajectories with $\kappa^2$'s which are sampled from a uniform distribution from $0$ to $1$. We fixed $\rho^2 = 0.1$ for all subjects. The subject-specific intercepts for albumin trajectories were randomly sampled from the Normal distribution $N(\mu = 5.0, \sigma^2 = 0.5)$. We simulate 9 to 12 longitudinal albumin values per subject. Using the simulated albumin trajectories, we generated survival times from the Weibull distribution in equation (\ref{eq:survModel}) that is of the following form for each of the proposed models
\begin{itemize}

	\item \textbf{Model I}:
\begin{eqnarray}\label{eq:SimSurvModel1}
T_i | \tau, \beta_1, X_i(t) & \sim & \textrm{Weibull}(\tau, \lambda_i = \beta^{(s)}_{i0} + \beta_1 X_i(t) ),
\end{eqnarray}

	\item \textbf{Model II}:
\begin{eqnarray}\label{eq:SimSurvModel2}
T_i | \tau, \beta_1, X_i(t) & \sim & \textrm{Weibull}(\tau, \lambda_i = \beta^{(s)}_{i0} + \beta_1 X_i(t) + \beta_2  X'_{AUC, i}(t) ),
\end{eqnarray}

	\item \textbf{Model III}:
\begin{eqnarray}\label{eq:SimSurvModel3}
T_i | \tau, \beta_1, X_i(t) & \sim & \textrm{Weibull}(\tau, \lambda_i = \beta^{(s)}_{i0} + \beta_1 Gender + \beta_2 {{\beta_0}^{(L)}} + \beta_3 {{\kappa_i}^2}).
\end{eqnarray}
\end{itemize}
The true values of the coefficients are set as follows
\begin{itemize}
	\item model I: $\beta_1 = -0.5$,
  \item model II: $\beta_1 = 0.5$, 
  \item model III: $\beta_1 = 0.5$, 
\end{itemize}
where in all simulations, $\beta^{(s)}_{i0}$ are simulated from a mixture of two Normal distributions of the form $\theta_{i} N(\mu = -1.5, \sigma^{2} = 1) + (1-\theta_{i}) N(\mu = 1.5, \sigma^{2} = 1)$, where $\theta_i$ is distributed Bernoulli with parameter $p = 0.5$.

Finally, the censoring times were sampled from a uniform distribution and independently from the simulated event times with an overall censoring rate of 20\%. 

All results are from 200 simulated datasets of size $n=300$ subjects each. For each dataset, we fit our proposed joint models with 10,000 draws where the first 5,000 considered as a burn-in period. Relatively diffuse priors were considered for all parameters. Details of the priors used in the simulations ass well as the results are as follow.

\subsection{Simulation Results for Model I}
In order to compare our proposed joint longitudinal-survival model that is capable of flexibly modeling longitudinal trajectories with simpler models with explicit functional assumptions on the longitudinal trajectories, we simulated longitudinal data once from quadratic polynomial longitudinal trajectory curves and another time from random non-linear curves. We then fit our joint model with a Gaussian Process longitudinal component as well as a joint model with the explicit assumption that the longitudinal trajectories are from a quadratic polynomial curve. As a comparison model, we also fit a two-stage Cox model where in stage one longitudinal data are modeled using our proposed Gaussian process longitudinal model and in the second stage, given the posterior mean parameters from the longitudinal fit, a Cox proportional hazard will fit the survival data. 

In particular, we generate synthetic longitudinal and survival data on 300 subjects, each with 9 to 12 within subject longitudinal albumin measures. Under the scenario where the longitudinal data are generated from quadratic polynomial longitudinal trajectories, we consider quadratic polynomial curves of the form 
\begin{align*}
X_{ij} = \beta^{(L)}_{0i} + \beta_{1i} t + \beta_{2} t^2 + \epsilon_{ij},
\end{align*}
where the true value of $\beta_{0i}$ are simulated from the Normal distribution $N(\mu = 5, \sigma^{2} = 1)$, $\beta_{1i}$ are simulated from the Normal distribution $N(\mu = -0.5, \sigma^{2} = 0.01)$, $\beta_{2}$ is set to be equal to -0.1, and finally $\epsilon_{ij}$ is the measurement error that is independent across measures and across subjects and are simulated from the Normal distribution $N(\mu = 0, \sigma = 0.1)$. 

Under the second scenario, longitudinal albumin values are generated from random non-linear curves. In particular, we generate random non-linear albumin trajectories that are realizations of a Gaussian process that are centered around the subject-specific random intercepts $\beta^{(L)}_{0i}$ that are generated from the Normal distribution $N(\mu = 5, \sigma^{2} = 1)$. We consider a Gaussian process with the squared exponential covariance function with the correlation length of $\rho^2 = 0.1$ and the subject-specific measures of volatility $\kappa^2_i$ that are generated from the uniform distribution $U(0, 1)$. 

For each simulation scenario, once longitudinal measures are generated, we generate survival data where survival times are distributed according to the Weibull distribution Weibull$(\tau, \lambda_i)$, where the shape parameter $\tau$ is set to 1.5 and $\lambda_i$, which is the log of the scale parameter of the Weibull distribution, is set to $\beta_{i0}^{(S)} + \beta_1 X_i(t)$, where $\beta_{i0}^{(S)}$ are generated from an equally weighted mixture of two Normal distributions of $N(\mu = -1.5, \sigma^{2} = 1)$ and $N(\mu = 1.5, \sigma^{2} = 1)$, $\beta_1$ is fixed to -0.5, and $X_i(t)$ is the longitudinal value for subject $i$ at time $t$ that is already simulated in the longitudinal step of the data simulation.

Our proposed joint longitudinal-survival model assumes the Normal prior $N(\mu = 5, \sigma^{2} = 4)$ on the random intercepts $\beta^i_0$, the log-Normal prior log-Normal$(-1, 2)$ on $\kappa^2_i$, the log-Normal prior log-Normal$(-1, 1)$ on $\sigma^2$, the log-Normal prior log-Normal$(0, 1)$ on $\tau$, the Normal prior $N(\mu = 0, \sigma^{2} = 25)$ on the survival shared intercept $\beta_0$, the Normal prior $N(\mu = 0, \sigma^{2} = 25)$ on the survival coefficient $\beta_1$, the Gamma prior $\Gamma(3, 3)$ on the concentration parameter of the Dirichlet distribution, and the Normal prior $N(\mu = 0, \sigma^{2} = 25)$ as the base distribution of the Dirichlet distribution. 

As the results in Table \ref{SimTableModel1Ch4} show, when data are simulated with a longitudinal trajectories that are quadratic polynomial curves, the joint polynomial model performs better in terms of estimating the albumin coefficient in the survival model with a smaller mean squared error compared to our proposed joint longitudinal-survival. In real world, however, the true functional forms of the trajectories of the biomarkers are not known. Under a general case where the biomarker trajectories can be any random non-linear curve (scenario 2), our proposed joint model outperforms the joint polynomial model. Further, our joint modeling framework that is capable of estimating differential subject-specific log baseline hazards provides significantly better coefficient estimates compared to the proportional hazard Cox model. Estimates under the Cox model are marginalized over all subjects and due to the non-collapsibility aspect of this model \citep{struthers1986,martinussen2013}, coefficient estimates shrink toward 0.
 
\begin{table}[ht]
\begin{center}
\scriptsize
\begin{tabular}{lcccccccccccc}
\hline
\multicolumn{1}{c}{Covariate of} &True Conditional& \multicolumn{3}{c}{Two-Stage Cox} && \multicolumn{3}{c}{Joint Polynomial Model} && \multicolumn{3}{c}{Joint Model}\\
\cline{3-5} \cline{7-9} \cline{11-13}
\multicolumn{1}{c}{Interest} & Estimand $$ & Mean & SD & MSE $$ && Mean & SD & MSE$$ && Mean & SD & MSE$$\\
\hline\\
\underline{Scenario 1}\\
~~~~~Albumin(t) & -0.5 & -0.273 & 0.056 & 0.119 && -0.495 & 0.019 & 0.003 && -0.441 & 0.105 & 0.012 \\
\\
\underline{Scenario 2}\\
~~~~~Albumin(t) & -0.5 & -0.258 & 0.080 & 0.125 && -0.380 & 0.080 & 0.034 && -0.462 & 0.110 & 0.010 \\
\hline
\end{tabular}
\normalsize
\caption{Model I Simulation results - joint longitudinal-survival data were generated under the simulation scenarios of one when longitudinal measures are sampled from the quadratic polynomial trajectories (scenario 1) and another scenario when longitudinal measures are sampled from random non-linear curves (scenario 2). Under each scenario, we fit three models of a joint longitudinal-survival model with the assumption that longitudinal trajectories are quadratic polynomial (Joint Polynomial Model), our proposed joint longitudinal-survival with a flexible Gaussian process longitudinal component (Joint Model), and a two-stage Cox proportional model with longitudinal trajectories with parameters that set to the posterior mean of a Gaussian process longitudinal model that is fit separately.}
\label{SimTableModel1Ch4}
\end{center}
\end{table}

\subsection{Simulation Results for Model II}
In model II, not only do we adjust for the albumin value at time $Y_i$, but we also adjust for a weighted average slope of albumin from time $\tau_1 = 0$ up until the time $\tau_2 = Y_i$, where $Y_i$ is either the event time for subject $i$ or is the time that the subject got censored. This new model differentiates between the risk of death for a patient whose albumin value is improving compared to another patient with the same albumin level whose albumin is deteriorating. In particular, we consider weighted average slope of albumin once under the weighting scheme of the form
\begin{align*}
Q(t) = \frac{1}{\tau_1 - \tau_0},
\end{align*}
and another time under the weighting scheme of 
\[
Q(t)= 
\begin{cases}
  1,& \text{if } t = T_i\\
  0,& \text{otherwise}.
\end{cases}
\]
The first weighting scheme leads to the area under the derivative curve. The second weighting scheme will result in a point-wise derivative of albumin at time $Y_i$.

We generate synthetic data for 300 subjects each with 9 to 12 longitudinal measurements where longitudinal albumin values are generated from a Gaussian process that is centered around the subject-specific random intercepts $\beta^{(L)}_{0i}$ which are generated from the Normal distribution $N(\mu = 5, \sigma^{2} = 1)$. We consider a Gaussian process with the squared exponential covariance function with the correlation length of $\rho^2 = 0.1$ and the subject-specific measures of volatility $\kappa^2_i$ that are generated from the uniform distribution $U(0, 1)$. Once longitudinal measures are generated, we generate survival data where survival times are distributed according to the Weibull distribution Weibull$(\tau, \lambda_i)$, where the shape parameter $\tau$ is set to 1.5 and $\lambda_i$, which is the log of the scale parameter in Weibull distribution, is set to $\beta_{i0}^{(S)} + \beta_1 X_i(t) + \beta_2 X'_{AUC, i}(t)$, where $\beta_{i0}^{(S)}$ are generated from an equally weighted mixture of two Normal distributions of $N(\mu = -1.5, \sigma^{2} = 1)$ and $N(\mu = 1.5, \sigma^{2} = 1)$, $\beta_1$ is fixed to 0.3, $\beta_2$ is fixed to 0.5, $X_i(t)$ is the longitudinal value for subject $i$ at time $t$ and $X'_{AUC, i}(t)$ is the average slope of albumin.

Our proposed joint longitudinal-survival model assumes the Normal prior $N(\mu = 5, \sigma^{2} = 4)$ on the random intercepts $\beta^i_0$, the log-Normal prior log-Normal$(-1, 2)$ on $\kappa^2_i$, the log-Normal prior log-Normal$(-1, 1)$ on $\sigma^2$, the log-Normal prior log-Normal$(0, 1)$ on $\tau$, the Normal prior $N(\mu = 0, \sigma^{2} = 25)$ on the survival shared intercept $\beta_0$, the Normal prior $N(\mu = 0, \sigma^{2} = 25)$ on the survival coefficient $\beta_1$, the Normal prior $N(\mu = 0, \sigma^{2} = 25)$ on the survival coefficient $\beta_2$, the Gamma prior $\Gamma(3, 3)$ on the concentration parameter of the Dirichlet distribution, and the Normal prior $N(\mu = 0, \sigma^{2} = 25)$ as the base distribution of the Dirichlet distribution. 

We fit our proposed joint longitudinal-survival model. As a comparison, we also fit a two-stage Cox model where the longitudinal curve of albumin and its derivative curve are estimated using hyper-parameters set as the posterior median of a Bayesian Gaussian Process model. As we can see from table \ref{SimTableModel2AUC}, our joint model provides closer estimates to the coefficient values with a smaller mean squared error compared with the two-stage Cox model. Our proposed model is capable of detecting differential subject-specific baseline hazards whereas the Cox model is not capable of differentiating between subjects and provides estimates that are marginalized across all subjects. Further, the simulation results show the capability of our method in detecting the true underlying longitudinal curves and the ability of our method on properly estimating the average derivative of those curves. 

\begin{table}[ht]
\begin{center}
\scriptsize
\begin{tabular}{lccccccccc}
\hline
\multicolumn{1}{c}{Covariate of} &True Conditional& \multicolumn{3}{c}{Two-Stage Cox} && \multicolumn{3}{c}{Joint Model}\\
\cline{3-5} \cline{7-9}
\multicolumn{1}{c}{Interest} & Estimand $$ & Mean & SD & MSE$$ && Mean & SD & MSE$$\\
\hline\\
\underline{Case 1 - Uniform Weights}\\
~~~~~Albumin(t) & 0.3 & 0.191 & 0.099 & 0.022 && 0.303 & 0.109 & 0.008\\
~~~~~Area under the derivative curve(t) & 0.5 & 0.346 & 0.179 & 0.053 && 0.449 & 0.188 & 0.030\\
\\
\underline{Case 2 - Point-Wise Weights}\\
~~~~~Albumin(t) & 0.3 & 0.142 & 0.095 & 0.033 && 0.261 & 0.104 & 0.009\\
~~~~~$\frac{d(Albumin(t))}{dt}$ & 0.5 & 0.412 & 0.123 & 0.022 && 0.477 & 0.152 & 0.013\\
\\
\hline
\end{tabular}
\normalsize
\caption{Model II simulation results - joint longitudinal-survival data were generated for 300 subjects each with 9 to 12 within subject measurements where longitudinal albumin values are generated from a Gaussian process that is centered around the subject-specific random intercepts $\beta^{(L)}_{0i}$ which are generated from the Normal distribution $N(\mu = 5, \sigma^{2} = 1)$. We consider a Gaussian process with the squared exponential covariance function with the correlation length of $\rho^2 = 0.1$ and the subject-specific measures of volatility $\kappa^2_i$ that are generated from the uniform distribution $U(0, 1)$. Once longitudinal measures are generated, we generate survival data where survival times are distributed according to the Weibull distribution Weibull$(\tau, \lambda_i)$, where the shape parameter $\tau$ is set to 1.5 and $\lambda_i$, which is the log of the scale parameter in Weibull distribution, is set to $\beta_{i0}^{(S)} + \beta_1 X_i(t) + \beta_2 X'_{AUC, i}(t)$, where $\beta_{i0}^{(S)}$ are generated from an equally weighted mixture of two Normal distributions of $N(\mu = -1.5, \sigma^{2} = 1)$ and $N(\mu = 1.5, \sigma^{2} = 1)$, $\beta_1$ is fixed to 0.3, $\beta_2$ is fixed to 0.5, $X_i(t)$ is the longitudinal value for subject $i$ at time $t$ and $X'_{AUC, i}(t)$ is the average slope of albumin. We fit our proposed joint longitudinal-survival model as well as a two-stage Cox proportional hazard model as a the comparison model.}
\label{SimTableModel2AUC}
\end{center}
\end{table}

\subsection{Simulation Results for Model III}
In model III, we test the association between the summary measures of the longitudinal biomarker trajectories and the survival outcomes. In particular, we consider the relation between the summary measures of subject-specific random intercept $\beta^{(L)}_{i0}$ and subject-specific measure of volatility $\kappa^2_i$ and survival times. 

We generate synthetic data for $N=300$ subjects each with 9 to 12 longitudinal measurements where longitudinal albumin values are generated from a Gaussian process that is centered around the subject-specific random intercepts $\beta^{(L)}_{0i}$ which are generated from the Normal distribution $N(\mu = 5, \sigma = 1)$. We consider a Gaussian process with the squared exponential covariance function with the correlation length of $\rho^2 = 0.1$ and the subject-specific measures of volatility $\kappa^2_i$ that are generated from the uniform distribution $U(0, 1)$. Once longitudinal measures are generated, we generate survival data where survival times are distributed according to the Weibull distribution Weibull$(\tau, \lambda_i)$, where the shape parameter $\tau$ is set to 1.5 and $\lambda_i$, which is the log of the scale parameter in Weibull distribution, is set to $\beta_{i0}^{(S)} + \beta_1 Age + \beta_2 {\beta_{i0}}^{(L)} + \beta_3 {\kappa_i^{2}}^{(L)}$, where $\beta_{i0}^{(S)}$ are generated from an equally weighted mixture of two Normal distributions of $N(\mu = -1.5, \sigma^{2} = 1)$ and $N(\mu = 1.5, \sigma^{2} = 1)$, $\beta_1$ is fixed to 0.5, $\beta_2$ is fixed to -0.3, $\beta_3$ is fixed to 0.7, $Age$ is a standardized covariate that is generated from the Normal distribution $N(\mu = 0, \sigma^{2} = 1)$, ${\beta_{i0}}^{(L)}$ is subject-specific random intercepts of the longitudinal trajectories, and ${\kappa_i^{2}}^{(L)}$ are subject specific measure of volatility of the longitudinal trajectories.

Our proposed joint longitudinal-survival model assumes the Normal prior $N(\mu = 5, \sigma^{2} = 4)$ on the random intercepts $\beta^i_0$, the log-Normal prior log-Normal$(-1, 2)$ on $\kappa^2_i$, the log-Normal prior log-Normal$(-1, 1)$ on $\sigma^2$, the log-Normal prior log-Normal$(0, 1)$ on $\tau$, the Normal prior $N(\mu = 0, \sigma^{2} = 25)$ on the survival shared intercept $\beta_0$, the Normal prior $N(\mu = 0, \sigma^{2} = 25)$ on the survival coefficient $\beta_1$, the Normal prior $N(\mu = 0, \sigma^{2} = 25)$ on the survival coefficient $\beta_2$, the Gamma prior $\Gamma(3, 3)$ on the concentration parameter of the Dirichlet distribution, and the Normal prior $N(\mu = 0, \sigma^{2} = 25)$ as the base distribution of the Dirichlet distribution. 

We fit our proposed joint survival-longitudinal model (model III) as well as a two-stage Cox proportional hazard model as a comparison model. The two-stage Cox model is a simple Cox proportional hazard model with covariate $\beta_{0i}^{(L)}$ and ${\kappa_i^2}^{(L)}$ that are posterior medians from a separate longitudinal Gaussian process model. As the results in Table \ref{SimTableModel3} show, our proposed joint model provides closer estimates to the true coefficients that also have significantly smaller mean squared error compared to the two-stage Cox model. Our proposed joint model is capable of detecting the differential subject-specific baseline hazards. Unlike our model, Cox model is blind to the subject-specific baseline hazards and hence, provides coefficient estimates that are marginalized over all subjects. These marginalized estimates from the Cox model shrink toward 0 as the Cox model with a multiplicative hazard function is non-collapsible. 

As one can see in the joint model results in Table \ref{SimTableModel3}, the coefficient estimate for ${\kappa^2}^{(L)}$ is not as close to the true coefficient value compared with other coefficient estimates. This is due to the fact that only 9 to 12 longitudinal measures per subject, there exists many plausible ${\kappa^2_i}^{(L)}$ values that flexibly characterize the trajectory of the measured albumin values. This additional variability in plausible ${\kappa^2_i}^{(L)}$ values has caused the coefficient estimate to shrink toward 0. Larger number of within subject longitudinal measures will provide more precision in estimating the true underlying ${\kappa^2_i}^{(L)}$ and will lead to a coefficient estimate closer to the true value. In order to confirm this fact, we simulated additional data once with 36 within subject measures and another time with 72 within subject measures. Table \ref{SimTableModel3Bonus} shows the results of fitting our proposed joint longitudinal-survival model to datasets that include subjects with 9 to 12 within subject measurements, to datasets with subjects with 36 within subject measurements, and to datasets with subjects with 72 within subject measurements. As the results show, with larger number of within subject measurements, coefficient estimate for ${\kappa^2_i}^{(L)}$ is closer to the true value. This is due to the fact that with larger number of within subject albumin measurements, there exists a stronger likelihood to estimate the subject-specific volatility measures $\kappa^2_i$, and hence, there is less uncertainty about the estimated value of volatility measures.

\begin{table}[ht]
\begin{center}
\scriptsize
\begin{tabular}{lccccccccc}
\hline
\multicolumn{1}{c}{Covariate of} &True Conditional& \multicolumn{3}{c}{Two-Stage Cox} && \multicolumn{3}{c}{Joint Model}\\
\cline{3-5} \cline{7-9}
\multicolumn{1}{c}{Interest} & Estimand $$ & Mean & SD & MSE$$ && Mean & SD & MSE$$\\
\hline\\
~~~~~Age (scaled) & 0.5 & 0.262 & 0.124 & 0.070 && 0.492 & 0.149 & 0.013\\
~~~~~Baseline Albumin (${\beta_{0i}}^{(L)}$) & -0.3 & -0.141 & 0.118 & 0.040 && -0.284 & 0.116 & 0.008\\
~~~~~${\kappa^2_i}^{(L)}$ & 0.7 & 0.414 & 0.212 & 0.127 && 0.595 & 0.271 & 0.042\\
\hline
\end{tabular}
\normalsize
\caption{Model III simulation results - joint longitudinal-survival data were generated for 300 subjects each with 9 to 12 longitudinal measurements where longitudinal albumin values are generated from a Gaussian process that is centered around the subject-specific random intercepts $\beta^{(L)}_{0i}$ which are generated from the Normal distribution $N(\mu = 5, \sigma^{2} = 1)$. We consider a Gaussian process with the squared exponential covariance function with the correlation length of $\rho^2 = 0.1$ and the subject-specific measures of volatility $\kappa^2_i$ that are generated from the uniform distribution $U(0, 1)$. Once longitudinal measures are generated, we generate survival data where survival times are distributed according to the Weibull distribution Weibull$(\tau, \lambda_i)$, where the shape parameter $\tau$ is set to 1.5 and $\lambda_i$, which is the log of the scale parameter in Weibull distribution, is set to $\beta_{i0}^{(S)} + \beta_1 Age + \beta_2 {\beta_{i0}}^{(L)} + \beta_3 {\kappa_i^{2}}^{(L)}$, where $\beta_{i0}^{(S)}$ are generated from an equally weighted mixture of two Normal distributions of $N(\mu = -1.5, \sigma^{2} = 1)$ and $N(\mu = 1.5, \sigma^{2} = 1)$, $\beta_1$ is fixed to 0.5, $\beta_2$ is fixed to -0.3, $\beta_3$ is fixed to 0.7, $Age$ is a standardized covariate that is generated from the Normal distribution $N(\mu = 0, \sigma^{2} = 1)$, ${\beta_{i0}}^{(L)}$ are subject-specific random intercepts of the longitudinal trajectories, and ${\kappa_i^{2}}^{(L)}$ are subject specific measures of volatility of the longitudinal trajectories. We fit our proposed joint longitudinal-survival model as well as a two-stage Cox proportional hazard model as a the comparison model.}
\label{SimTableModel3}
\end{center}
\end{table}

\begin{table}[ht]
\begin{center}
\scriptsize
\begin{tabular}{lcccccccccccc}
\hline
\multicolumn{1}{c}{Covariate of} &True Conditional& \multicolumn{3}{c}{Joint Model ($l_i = 12$)} && \multicolumn{3}{c}{Joint Model ($l_i = 36$)} && \multicolumn{3}{c}{Joint Model ($l_i = 72$)}\\
\cline{3-5} \cline{7-9} \cline{11-13}
\multicolumn{1}{c}{Interest} & Estimand $$ & Mean & SD & MSE$$ && Mean & SD & MSE$$ && Mean & SD & MSE$$\\
\hline\\
~~~~~Age (scaled) & 0.5 & 0.492 & 0.149 & 0.013 && 0.493 & 0.144 & 0.015 && 0.495 & 0.145 & 0.016\\
~~~~~${\beta_{i0}}^{(L)}$ & -0.3 & -0.284 & 0.116 & 0.008 && -0.308 & 0.116 & 0.006 && -0.295 & 0.115 & 0.007\\
~~~~~${\kappa^2_i}^{(L)}$ & 0.7 & 0.595 & 0.271 & 0.042 && 0.639 & 0.284 & 0.043 && 0.651 & 0.293 & 0.039\\
\hline
\end{tabular}
\normalsize
\caption{Model III simulation results with datasets with $l_i = 36$ and $l_i = 72$ within subject measurements. In order to test the sensitivity of the ${\kappa^2}^{(L)}$ coefficient estimate to the number of within subject measurements, $l_i$, we simulated joint longitudinal-survival data once when each subject has 36 within subject measurements and another time when each subject has 72 within subject measurements. Under each scenario, we simulated 200 datasets each with 300 subjects. Other simulation parameters remained the same as the simulation parameters used in Table \ref{SimTableModel3}. This means, we simulated longitudinal data from Gaussian process that is centered around the subject-specific random intercepts $\beta^{(L)}_{0i}$ which are generated from the Normal distribution $N(\mu = 5, \sigma^{2} = 1)$. We consider a Gaussian process with the squared exponential covariance function with the correlation length of $\rho^2 = 0.1$ and the subject-specific measures of volatility $\kappa^2_i$ that are generated from the uniform distribution $U(0, 1)$. Once longitudinal measures are generated, we generate survival data where survival times are distributed according to the Weibull distribution $Weibull(\tau, \lambda_i)$, where the shape parameter $\tau$ is set to 1.5 and $\lambda_i$, which is the log of the scale parameter in Weibull distribution, is set to $\beta_{i0}^{(S)} + \beta_1 Age + \beta_2 {\beta_{i0}}^{(L)} + \beta_3 {\kappa_i^{2}}^{(L)}$, where $\beta_{i0}^{(S)}$ are generated from an equally weighted mixture of two Normal distributions of $N(\mu = -1.5, \sigma^{2} = 1)$ and $N(\mu = 1.5, \sigma^{2} = 1)$, $\beta_1$ is fixed to 0.5, $\beta_2$ is fixed to -0.3, $\beta_3$ is fixed to 0.7, $Age$ is a standardized covariate that is generated from the Normal distribution $N(\mu = 0, \sigma^{2} = 1)$, ${\beta_{i0}}^{(L)}$ are subject-specific random intercepts of the longitudinal trajectories, and ${\kappa_i^{2}}^{(L)}$ are subject specific measures of volatility of the longitudinal trajectories.}
\label{SimTableModel3Bonus}
\end{center}
\end{table}


\section{Application of the Proposed Joint Models to ESRD}
\label{uniJointReal}

In this section, we apply our proposed joint longitudinal-survival models to data on $n = 1,112$ end stage renal disease patients participating in the Dialysis Morbidity and Mortality Studies (DMMS) nutritional study that is obtained from the United States Renal Data System. For every participating patient in the study, up to 12 albumin measurements were taken uniformly over two years of followup. The presented analyses are restricted to only the patients who had at least nine albumin measurements in order to provide sufficient data for modeling the trajectory and the volatility of albumin. The censoring rate in the data is at 43\% over a maximal follow-up time of 4.5 years. 

Using the same data, \cite{fung02} showed that both baseline albumin level and the slope of albumin over time are significant predictors of mortality among ESRD patients. While our models are capable of replicating Fung et al's findings, our models are also capable of:
\begin{itemize}
	\item model 1: testing the association between albumin value at the time of death and the risk of death
    \item model 2: testing the association between albumin value and an average derivative of albumin up until time t and the risk of death.
    \item model 3: Testing the association between risk of mortality and the two summary measures of the baseline and the volatility of albumin measures 
\end{itemize}
In order to adjust for other potential confounding factors, our proposed models also include patient's age, gender, race, smoking status, diabetes, an indicator of whether the patient appeared malnourished at baseline, BMI at baseline, baseline cholesterol, and baseline systolic blood pressure. The adjusted covariates are consistent with those originally presented in \cite{fung02}.

Table \ref{AppRslt1} and Table \ref{AppRslt2auc} provide the results of fitting our proposed model I, Model II, and Model III to the USRDS data. All joint models were run for 10,000 posterior samples where the initial 5,000 samples are discarded as burn-in samples. 


    
    

\subsection{Results for Model I}
\normalsize
We fit our proposed joint model I to the data. As a comparison model, we also fit as last-observation carried forward (LOCF) Cox model. Table \ref{AppRslt1} shows the estimated coefficients from both models. Between the two models, the estimated relative risk associated with all time-invariant baseline survival covariates are similar between the two models. However, the relative risk associated with every one unit decrement in serum albumin is much larger under our proposed joint model compared to the last-observation carried forward Cox model. This is quite expected as our model is capable of estimating subject-specific albumin trajectories over time and is capable of accurately testing the association between albumin value at time of death and risk of death. Unlike our model,the LOCF Cox model uses the most recent albumin measure which in reality might be quite different than the albumin value at the time of death. In both models, albumin is identified as a significant risk factor of mortality. In particular, based on the results from our proposed joint Model I, it is estimated that every 1 $g/dL$ decrement in albumin is associated with a 4.5 times higher risk of death. 

\begin{table}[t!]
\begin{center}
\vspace{12pt}
\scriptsize  
\begin{tabular}{lrrcccc}
\hline
                              &       &      &  \multicolumn{1}{c}{LOCF Cox Model}      & &    \multicolumn{1}{c}{Joint Model}           \\
 \cline{4-4} \cline{6-7}
                              & No. of& No. of& \multicolumn{1}{c}{Relative Risk}   & & \multicolumn{1}{c}{Relative Risk}       \\
Covariates                    & Cases &  Deaths    &  (95\% CI) & P-Value &(95\% CR)     &  \\
\hline
Age (10y)                          &1,112&630&1.44 (1.35-1.53)&$<$.001&1.45 (1.36,1.55)&\\ 
Sex                                &     &   &                &       &                &       \\ 
\hspace{0.5cm} Men                 &  560&312&1.0             &       & 1.0            &       \\ 
\hspace{0.5cm} Women               &  552&318&0.96 (0.81,1.13)&0.60  &0.97 (0.82,1.16)& \\ 
Race                               &     &   &                &       &                &       \\
\hspace{0.5cm} White               &  542&350&1.0             &       & 1.0            &       \\
\hspace{0.5cm} Black               &  482&243&0.81 (0.68,0.96)&0.01  &0.79 (0.67,0.94)& \\
\hspace{0.5cm} Other               &  88& 37&0.52 (0.37,0.74)&$<$.001  &0.49 (0.34,0.69)&\\
Smoking                            &     &   &                &       &                &       \\
\hspace{0.5cm} Nonsmoker           &  645&337&1.0             &       & 1.0            &       \\
\hspace{0.5cm} Former              &  307&197&1.17 (0.98,1.41)&0.09  &1.20 (0.99,1.44)& \\
\hspace{0.5cm} Current             &  160&96&1.52 (1.19,1.94)&$<$.001  &1.53 (1.21,1.95)&\\
Diabetes                           &     &   &                &       &                &       \\
\hspace{0.5cm} No                  &  716&363&1.0             &       & 1.0            &       \\
\hspace{0.5cm} Yes                 &  396&267&1.66 (1.40,1.97)&$<$.001&1.69 (1.43,2.00)&\\
Undernourished                     &     &   &                &       &                &       \\
\hspace{0.5cm} No                  &958&517&1.0             &       & 1.0            &       \\
\hspace{0.5cm} Yes                 &  154&113&1.39 (1.12,1.72)&0.003  &1.35 (1.08,1.66)&\\
BMI (per-5 kg/m$^2$                &     &   &                &       &                &       \\
\hspace{0.85cm} decrement)         &1,112&630&1.08 (1.00,1.17)&0.07  &1.08 (1.00,1.17)&\\
Cholesterol (per 20                &     &   &                &       &                &       \\
\hspace{0.85cm} mg/dL)             &1,112&630&0.97 (0.93,1.00)&0.08  &0.96 (0.93,1.00)& \\
Systolic blood pressure            &     &   &                &       &                &       \\
\hspace{0.85cm} (per 10mm Hg)      &1,112&630&0.98 (0.95,1.02)&0.38  &0.98 (0.95,1.02)&\\
Serum albumin(t) (1-g/dL decrement)    &1,112&630&2.48 (2.00,3.07)&  $<$0.001 &4.54 (3.03,5.55)&\\
\hline
\end{tabular}
\caption{Estimated Relative Risk and corresponding 95\% credible region from our proposed joint model where we adjust for time-dependent albumin value that is imputed from the longitudinal component of the model. We also fit a last-observation carried forward Cox proportional hazards model with last albumin value carried forward where we report coefficients estimates, 95\% confidence interval, and p-value for the estimated coefficients. In both models, we adjust for potential confounding factors as reported by \cite{fung02}.}
\label{AppRslt1}
\end{center}
\end{table}
\normalsize
\subsection{Results for Model II}
Other than the albumin value at the time of death, the average slope of albumin over time might also be a risk factor of mortality in end-stage renal disease patients. In our proposed joint Model II, we also adjust for the area under the derivative curve of the albumin trajectory from the time that the follow-up starts until the survival time which is either the time of death or the censoring time. Table \ref{AppRslt2auc} shows the results from our proposed model. Based on the results, every one g/dL decrement in albumin is associated with 3.95 times higher risk of death. Also, higher average slope of albumin, that is every 1 g/dL/month increase in the average slope, is associated with 2.3 times higher risk of death. This is consistent with \cite{fung02} results on the association between the slope of albumin and the risk of death. Our proposed method is also capable of adjusting for the local effect of the slope of albumin. For instance, instead of averaging the slope of the follow-up time, one may only integrate over the 6 months prior to the time of death. 

\subsection{Results for Model III}
\begin{figure}[!htb]
\includegraphics[width=\linewidth]{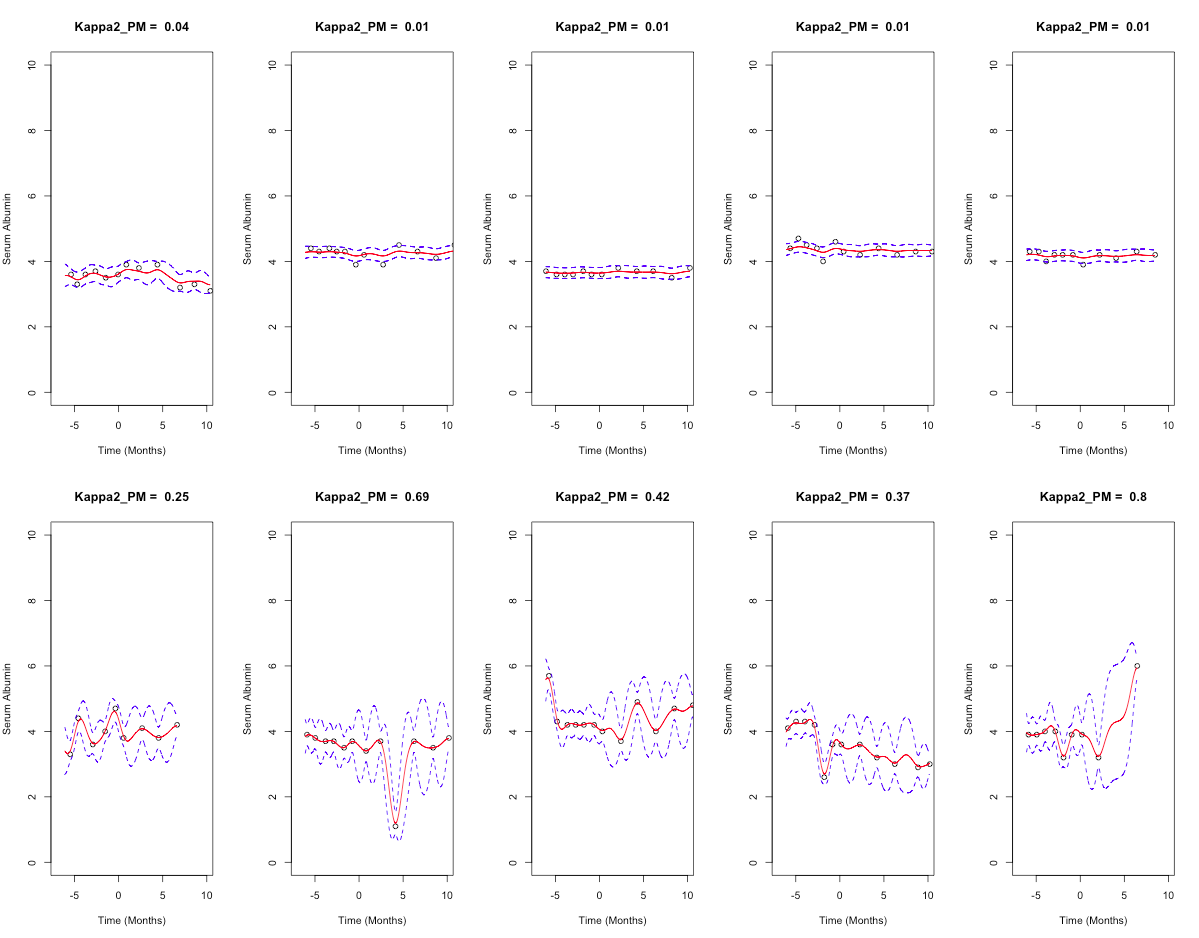}
\caption{Actual longitudinal albumin trajectories of 10 randomly selected individuals with end-stage renal disease that were selected from the USRDS data. Hollow circles are the actual measured albumin values, red lines are the posterior median fitted curves from our proposed Model III, and the dashed blue lines are the corresponding 95\% posterior prediction intervals for the fitted trajectories. The title of each plot shows the posterior median of the volatility measure $\kappa^2$ for the subject whose albumin measures are shown in the plot.}
\label{ActualUSRDSLong}
\end{figure}

\cite{fung02} showed that the baseline albumin and the slope of albumin over time are two independent risk factors of mortality among the end-stage renal disease patients. It is quite natural to hypothesize that the volatility of albumin could also be a risk factor of mortality among these patients. In our proposed joint longitudinal-survival Model III, we consider two summary measures of the trajectories of the longitudinal albumin values, one the baseline albumin measures (${\beta_{0i}}^{(L)}$), and another the subject-specific volatility measure of albumin (${\kappa^2_i}^{(L)}$). Table \ref{AppRslt2auc} also shows the results from our proposed Model III. The results from our model confirms that the baseline serum albumin is a risk factor of mortality. Further, the results from our model indicate that the volatility of albumin is also a significant risk factor of mortality, where every one unit increase in $\kappa^2$, which indicates a higher volatility, is associated with 1.2 times higher risk of death. Figure \ref{ActualUSRDSLong} shows albumin trajectories of 10 randomly sampled individuals. 

\begin{table}[t!]
\begin{center}
\vspace{12pt}
\footnotesize
\begin{tabular}{lrrcc}
\hline
                              &       &      &  \multicolumn{1}{c}{Joint Model}\\
 \cline{4-4}
                              & No. of& No. of& \multicolumn{1}{c}{Relative Risk}\\
Covariates                    & Cases &  Deaths    &  (95\% CI)     &\\
\hline\\
\underline{Model 2}\\
Serum Albumin(t) (1-g/dL decrement)    &1,112&630&3.95 (3.18,4.71)&\\
Average Derivative of Serum Albumin$^1$ (1-g/dL/month decrement) &1,112&630&2.33 (1.40,3.73)&\\
\\
\underline{Model 3}\\
Baseline Albumin$({\beta_{0i}}^{(L)})$ (1-g/dL decrement)    &1,112&630&5.54 (4.19,6.94)&\\
${\kappa^2_i}^{(L)}$ (increase in volatility)$^2$    &1,112&630&1.23 (1.02,1.41)&\\
\hline
\multicolumn{4}{l}{1 : One may only consider the local effect of average serum albumin slope by computing the area under the}\\
\multicolumn{4}{l}{~~~~~derivative from 6 months prior to death up until the time of death.}\\
\multicolumn{4}{l}{2 : In a similar model, we adjusted for $\kappa^2$ values as a categorical variable with a cut point equal to the }\\
\multicolumn{4}{l}{~~~~~posterior mean of all $\kappa^2$ values (0.1) and we got a similar estimate relative risk (1.21).}\\
\end{tabular}
\caption{Model II and Model III results that show the estimated relative risk and corresponding confidence intervals from our proposed joint model II and model III. Potential confounding factors, as reported by \cite{fung02}, were also adjusted in the model but have been removed from the tables for brevity. Our proposed Model II is capable of testing the association between albumin values at the time of death as well as the average derivative of the subject-specific albumin trajectories from the time the follow up time starts up until the death or the censoring time. Our proposed Model III tests the association risk of mortality and two albumin trajectory summary measures of the subject-specific random intercepts (${\beta_{0i}}^{(L)}$) and the subject specific volatility measures (${\kappa^2_i}^{(L)}$).}
\label{AppRslt2auc}
\end{center}
\end{table}

\section{Discussion}\label{uniJointDisc}
Monitoring the health of patients often involves recording risk factors over time. In such situations, it is essential to evaluate the association between those longitudinal measurements and survival outcome. To this end, joint longitudinal-survival models provide an efficient inferential framework. 

We proposed a joint longitudinal-survival framework that avoids some of the restrictive assumptions commonly used in the existing models. Further, our methods propose a stronger link between longitudinal and survival data through an introduction of new ways of adjusting for the biomarker value at time $t$, adjusting for the average derivative of the biomarker over time, and moving beyond the first-order trend and accounting for volatility of biomarker measures over time.  

Our proposed method can be considered as an extension of the joint model proposed by \cite{brown2003} in that we use the same  idea of dividing the joint likelihood into a marginal longitudinal likelihood and conditional survival likelihood. However, instead of fitting quadratic trajectories, we use a flexible longitudinal model based on the Gaussian processes. Further, for the survival outcome, instead of assuming a piecewise exponential model, we use a flexible survival model by incorporating the Dirichlet process mixture of Weibull distributions. Our proposed modeling framework is capable of modeling additional summary measures of longitudinally measured biomarkers and relating them to the survival outcome in a time-dependent fashion. 

Despite its flexibility and novelty, our approach has some limitations. By using th Bayesian non-parameteric Dirichlet process and the Gaussian process techniques, while we provide a flexible modeling framework that avoids common distributional assumptions, however, these techniques are generally not scalable when the number of subjects and the number of within subject measurements increase. Furthermore, the survival component of our model still relies on the proportional hazard assumption. In future, our modeling framework can be extended to include a more general non-proportional hazard survival models that can also include time-dependent coefficients inside the survival model. By using some alternatives to the common MCMC techniques, including parallel-MCMC methods and variational methods, our method can become more computationally efficient and scalable for larger datasets.

Often times in monitoring the health of patients, multiple longitudinal risk factors are measured. One can use our proposed modeling framework in this paper in order to build a joint longitudinal-survival model with multiple longitudinal processes each process modeled independently from other longitudinal processes. In reality, however, one expects that patients longitudinal risk factors to be correlated. A methodology that is capable of modeling multiple biomarkers simultaneously by taking the correlation between biomarkers into account can be beneficial.  

\section{Acknowledgements}
Babak Shahbaba was supported by the NIH grant R01 AI107034. 








\clearpage
\bibliographystyle{jasa}

\bibliography{refs}

\end{document}